\documentclass[11pt,preprint]{aastex}

\newcommand{\wisk}[1]{\ifmmode{#1}\else{$#1$}\fi}

\slugcomment{KSUPT-99/6 \hspace{0.5truecm} September 1999}

\begin{document}

\title{Supernovae Ia Constraints on a Time-Variable Cosmological ``Constant"}

\author{Silviu Podariu and Bharat Ratra}

\affil{Department of Physics, Kansas State University, Manhattan, KS 66506.}

\begin{abstract}
The energy density of a scalar field $\phi$ with potential $V(\phi) \propto 
\phi^{-\alpha}$, $\alpha > 0$, behaves like a time-variable cosmological 
constant that could contribute significantly to the present energy density.
Predictions of this spatially-flat model are compared to recent Type Ia
supernovae apparent magnitude versus redshift data. A large region of model
parameter space is consistent with current observations. (These constraints are based on the exact scalar field model equations of motion, not on the widely 
used time-independent equation of state fluid approximation equations of 
motion.) We examine 
the consequences of also incorporating constraints from recent measurements of 
the Hubble parameter and the age of the universe in the constant and 
time-variable cosmological constant models. We also study the effect of using a 
non-informative prior for the density parameter. 
\end{abstract}

% Keywords

\section{Introduction}

Current observations are more consistent with low-density cold dark matter
(CDM) dominated cosmogonies. For recent discussions see, e.g., Ratra et al. 
(1999a), Kravtsov \& Klypin (1999), Doroshkevich et al. (1999), 
Colley et al. (2000), Freudling et al. (1999), Sahni \& Starobinsky 
(1999), Bahcall et al. (1999), and Donahue \& Voit (1999). The simplest 
low-density CDM models have either 
flat spatial hypersurfaces and a constant or time-variable cosmological 
``constant" $\Lambda$ (see, e.g., Peebles 1984; Peebles \& Ratra 1988, hereafter
PR; Efstathiou, Sutherland, \& Maddox 1990; Stompor, G\'orski, \& Banday 1995;
Caldwell, Dave, \& Steinhardt 1998), or open spatial hypersurfaces and no 
$\Lambda$ (see, e.g., Gott 1982, 1997; Ratra \& Peebles 1994, 1995; 
G\'orski et al. 1998). 

While these models are consistent with most recent observations, there are
notable exceptions. For instance, recent applications of the apparent magnitude 
versus redshift test based on Type Ia supernovae favor a non-zero $\Lambda$
(see, e.g., Riess et al. 1998, hereafter R98; Perlmutter et al. 1999a, hereafter
P99), although not with great statistical significance (Drell, Loredo, \&
Wasserman 1999).  

On the other hand, the open model is favored by:
\begin{itemize}
\item Measurements of the Hubble parameter $H_0$ (= $100h\ {\rm km}\ 
{\rm s}^{-1}\ {\rm Mpc}^{-1}$) which indicate $h = 0.65 \pm 0.1$ 
at 2 $\sigma$ (see, e.g., Suntzeff et al. 1999; Biggs et al. 1999; Madore
et al. 1999), and measurements of the age of the universe which indicate 
$t_0 = 12 
\pm 2.5$ Gyr at 2 $\sigma$ (see, e.g., Reid 1997; Gratton et al. 1997; 
Chaboyer et al. 1998). This is because the resulting central $H_0 t_0$ value is
consistent with a low-density open model with nonrelativistic-matter 
density parameter $\Omega_0 
\approx 0.35$ and requires a rather large $\Omega_0 \approx 0.6$ in the 
flat-constant-$\Lambda$ case.
\item Analyses of the rate of gravitational lensing of quasars and radio
sources by foreground galaxies which require a large $\Omega_0 \geq 0.38$ at 
2 $\sigma$ in the flat-constant-$\Lambda$ model (see, e.g., Falco, Kochanek, 
\& Mu\~noz 1998).
\item Analyses of the number of large arcs formed by strong gravitational
lensing by clusters (Bartelmann et al. 1998; Meneghetti et al. 1999).
\item The need for mild anti-biasing to accommodate the excessive intermediate- 
and small-scale power predicted in the DMR normalized flat-constant-$\Lambda$ 
CDM model with a scale-invariant spectrum (see, e.g., Cole et al. 1997).
\end{itemize}

While the time-variable $\Lambda$ model has not yet been studied to the same 
extent as the open and flat-$\Lambda$ cases, it is likely that it can 
be reconciled with most of these observations (see, e.g., PR; Ratra \&
Quillen 1992, hereafter RQ; Frieman \& Waga 1998; Perlmutter, Turner, \& 
White 1999b; Wang et al. 1999; Efstathiou 1999).

We emphasize that most of these observational indications are tentative
and certainly not definitive. This is particularly true for constraints
derived from a $\chi^2$ comparison between model predictions and cosmic
microwave background (CMB) anisotropy measurements (see, e.g., Ganga, Ratra,
\& Sugiyama 1996; Lineweaver \& Barbosa 1998; Baker et al. 1999; Rocha 1999), see discussions in Bond, Jaffe, \& Knox (1998) and Ratra
et al. (1999b). More reliable constraints follow from models-based 
maximum likelihood analyses of the CMB anisotropy data (see, e.g., 
G\'orski et al. 1995; Yamamoto \& Bunn 1996; Ganga et al. 1998; 
Ratra et al. 1999b; Rocha et al. 1999), which make use of all the 
information in the CMB anisotropy data and are based on fewer approximations.
However, this technique has not yet been used to analyze enough data sets 
to provide robust statistical constraints. Kamionkowski \& Kosowsky (1999)
review what might be expected from the CMB anisotropy in the near future.

In this paper we examine constraints on a constant and time-variable 
$\Lambda$ that follow from recent Type Ia supernovae apparent
magnitude versus redshift data and recent measurements of $H_0$ and $t_0$.
We focus here on the favored scalar field ($\phi$) model for a time-variable 
$\Lambda$ (PR; Ratra \& Peebles 1988, hereafter RP), in which the scalar
field potential $V(\phi) \propto \phi^{-\alpha}$, $\alpha > 0$, at low
redshift\footnote{
Other potentials have been considered, e.g., an exponential potential (see,
e.g., Lucchin \& Matarrese 1985; RP; Ratra 1992; Wetterich 1995; Ferreira \& 
Joyce 1998), or one that gives rise to an ultra-light pseudo-Nambu-Goldstone
boson (see, e.g., Frieman et al. 1995; Frieman \& Waga 1998), but such
models are either inconsistent with observations or do not share
the more promising features of the inverse power law potential model.}.
This scalar field could have played the role of the inflaton at much higher
redshift, with the potential $V(\phi)$ dropping to a non-zero value at the end 
of inflation and then decaying more slowly with increasing $\phi$ (PR; RP).
See Peebles \& Vilenkin (1999), Perrotta \& Baccigalupi (1999) and 
Giovannini (1999) for a specific model and observational consequences of
this scenario. A potential $\propto \phi^{-\alpha}$ 
could arise in a number of high energy particle physics models, see, e.g., 
Bin\'etruy (1999), Kim (1999), Barr (1999), Choi (1999), Banks, Dine, \&
Nelson (1999), Brax \& Martin (1999), Masiero, Pietroni, \& Rosati (1999), 
and Bento \& Bertolami (1999) for specific examples. It is conceivable that
such a setting might provide an explanation for the needed form of the 
potential, as well as for the needed very weak coupling of $\phi$ to other 
fields (RP; Carroll 1998;
Kolda \& Lyth 1999, but see Periwal 1999 and Garriga, Livio, \& Vilenkin
1999 for other possible explanations for the needed present value of
$\Lambda$). 

A scalar field is mathematically equivalent to a fluid with a time-dependent
speed of sound (Ratra 1991). This equivalence may be used to show that
a scalar field with potential $V(\phi) \propto \phi^{-\alpha}$, $\alpha > 0$,
acts like a fluid with negative pressure and that the $\phi$ energy density
behaves like a cosmological constant that decreases with time. This energy
density could come to dominate at low redshift and thus help reconcile low
dynamical estimates of the mean mass density with a spatially-flat cosmological
model. Alternate mechanisms that also rely on negative pressure to achieve
this result have been discussed (see, e.g., \"Ozer \& Taha 1986; Freese et al. 
1987; Turner \& White 1997; Chiba, Sugiyama, \& Nakamura 1997;
\"Ozer 1999; Waga \& Miceli 1999; Overduin 1999; Bucher \& Spergel 1999).
However, these mechanisms are either inconsistent or do not share a very 
appealing feature of the scalar
field models. For some of the scalar field potentials mentioned above, the
solution of the equations of motion is an attractive time-dependent fixed 
point (RP; PR; Wetterich 1995; Ferreira \& Joyce 1998; Copeland, Liddle, \& 
Wands 1998; Zlatev, Wang, \& Steinhardt 1999; Liddle \& Scherrer 1999;
Santiago \& Silbergleit 1998; Uzan 1999). This means that for a wide range
of initial conditions the scalar field $\phi$ evolves in a manner that ensures
that the cosmological constant contributes a reasonable energy density at
low redshift. Of course, this does not resolve the (quantum-mechanical)
cosmological constant problem. 

There have been many studies of the constraints placed on a time-variable
$\Lambda$ from Type Ia supernovae apparent magnitude versus redshift data (see, 
e.g., Turner \& White 1997; Frieman \& Waga 1998; Garnavich et al. 1998; 
Hu et al. 1999; Cooray 1999;
Waga \& Miceli 1999; P99; Perlmutter et al. 1999b; Wang et al. 1999; 
Efstathiou 1999). However, except for the analysis of Frieman \& Waga (1998),
who use the earlier Perlmutter et al. (1997) data, these analyses have mostly
made use of the time-independent equation of state fluid approximation to the
scalar field model for a time-variable $\Lambda$ (Perlmutter et al. 1999b and 
Efstathiou 1999 do go beyond the  time-independent equation of state 
approximation by also approximating the time dependence of the equation of 
state,
however, they do not analyze the exact scalar field model)\footnote{
A similar criticism holds for most analyses of the constraints on a 
time-variable $\Lambda$ from gravitational lensing statistics (see, e.g., 
Bloomfield Torres \& Waga 1996; Jain et al. 1998; Cooray 1999; Waga \& Miceli
1999; Wang et al. 1999), with the exception of the analysis of RQ.}.
A major purpose of this paper is to use the newer supernovae data of R98 
and P99 to derive constraints on a time-variable $\Lambda$ without making
use of the time-independent equation of state fluid approximation to the
scalar field model.

In the analyses here we use the most recent R98 and P99 data to place 
constraints on a constant and time-variable $\Lambda$. We note, however,
that this is a young field and insufficient understanding of a number of
astrophysical processes and effects (the mechanism responsible for the 
supernova, evolution, environmental effects, intergalactic dust, etc.) 
could render this a premature undertaking (see, e.g., H\"oflich,
Wheeler, \& Thielemann 1998; Aguirre 1999; Drell et al. 1999;
Umeda et al. 1999; Riess et al. 1999; Wang 1999). Furthermore, other 
cosmological explanations (time-variable gravitational or fine structure
``constants") could be consistent with the data (see, e.g., Amendola, 
Corasaniti, \& Occhionero 1999; Barrow \& Magueijo 1999; Garc\'{\i}a-Berro 
et al. 1999).   

In addition to analyzing just the supernovae data, we also incorporate 
constraints from recent measurements of $H_0$ and $t_0$ and derive a 
combined likelihood function which we use to constrain both a constant
and a time-variable $\Lambda$. We also examine the effect on the model-parameter
constraints caused by varying the prior for $\Omega_0$.

We emphasize that the tests examined in this paper are not sensitive to the
spectrum of inhomogeneities in the models considered. This spectrum (possibly
generated by quantum mechanical zero-point fluctuations during inflation,
see, e.g., Fischler, Ratra, \& Susskind 1985) is relevant for some of the 
other tests mentioned above (CMB anisotropy, anti-biasing, etc.). 

In $\S$2 we summarize the computations. Results are presented and discussed 
in $\S$3. Conclusions are given in $\S$4.

\section{Computation}

We examine three sets of supernovae apparent magnitudes. We use the MLCS
magnitudes for the R98 data, both including and excluding the unclassified
SN 1997ck at $z = 0.97$ (with 50 and 49 SNe Ia, respectively). The third
set are the corrected/effective stretch factor magnitudes for the 54 Fit
C SNe Ia of P99. In all three cases we assume that the measured magnitudes
are independent. We also assume that SNe Ia at high and low $z$ are not
significantly different (Drell et al. 1999, Riess et al. 1999, and Wang
1999 consider the possibility and consequences of evolution), and that
intergalactic dust has a negligible effect (Aguirre 1999 considers the 
effects of dust). 

Our analysis of the R98 data is similar to that described in their $\S$4, with
a few minor modifications. For the time-independent $\Lambda$ model we compute
the likelihood function $L(\Omega_0, \Omega_\Lambda, H_0)$ for a range of
$\Omega_0$ spanning the interval 0 to 2.5 in steps of 0.1, for a range of
$\Omega_\Lambda$ spanning the interval $-1$ to 3 in steps of 0.1, and for a
range of $H_0$ spanning the interval 50 to 80 km s$^{-1}$ Mpc$^{-1}$ in 
steps of 0.5 km s$^{-1}$ Mpc$^{-1}$.

Our analysis of the P99 data is similar to theirs, with some
modifications. Specifically, we work with their corrected/effective 
magnitudes and so need only compute a three-dimensional likelihood 
function $L(\Omega_0, \Omega_\Lambda, {\cal M}_B)$. Here ${\cal M}_B$ is
their ``Hubble-constant-free" B band absolute magnitude (see P99) related
to $H_0$ by ${\cal M}_B = -19.46 - 5{\rm log}_{10} H_0 + 25$ (determined 
by us from results in P99). For the time-independent $\Lambda$ model we compute
this likelihood function for a range of $\Omega_0$ spanning the interval 0 to 
3 in steps of 0.1, for a range of $\Omega_\Lambda$ spanning the interval $-1.5$ 
to 3 in steps of 0.1, and for a range of ${\cal M}_B$ spanning the interval
$-3.95$ to $-2.95$ in steps of 0.05 (which corresponds to about the same 
interval in $H_0$ used in our analyses of the R98 data). Note that in our 
P99 time-independent $\Lambda$ model plots we do not show the likelihood 
for the whole $\Omega_0-\Omega_\Lambda$ region over which we have computed it.

The spatially-flat time-variable $\Lambda$ model we consider is the scalar
field model with potential $V(\phi) \propto \phi^{-\alpha}$, $\alpha > 0$. When 
$\alpha \rightarrow 0$ the model tends to the flat-constant-$\Lambda$ case and 
when $\alpha \rightarrow \infty$ it approaches the Einstein-de Sitter model.
It is discussed in detail in PR, RP, and RQ, and we derive the predicted 
distance moduli for the SNe using expressions given in these papers.

As shown in these papers, the scalar field behaves like a fluid with a 
constant (but different) equation of state in each epoch of the model.
For instance, in the CDM and baryon dominated epoch of the model, it obeys
the equation of state $p_\phi = w_\phi \rho_\phi$ (relating the pressure
and energy density of the scalar field), where
\begin{equation}
     w_\phi = -{2 \over \alpha + 2} ,
\end{equation}
see, e.g., RQ eq. (2) or see Zlatev et al. (1999) for a more recent derivation.
We shall also have need for an average equation of state parameter,
used by Perlmutter et al. (1999b) and Wang et al. (1999),
\begin{equation}
     w_{\rm eff} = {\int^{a_0}_0 da \, \Omega_\phi (a) w_\phi (a)
                    \over \int^{a_0}_0 da \, \Omega_\phi (a) }
\end{equation}
where $\Omega_\phi$ is the scalar field density parameter and $a$ is
the scale factor (with $a_0$ being the present value).

In this model, for the R98 data we compute the likelihood function
$L(\Omega_0, \alpha, H_0)$, and for the P99 data the likelihood 
function $L(\Omega_0, \alpha, {\cal M}_B)$. In both cases we 
evaluate the likelihood function for a range of
$\Omega_0$ spanning the interval 0.05 to 0.95 in steps of 0.025, for a range of
$\alpha$ spanning the interval 0 to 8 in steps of 0.5, and for 
the same range of $H_0$ or ${\cal M}_B$ as in the time-independent $\Lambda$
cases discussed above.

We marginalize these three-dimensional likelihood functions by integrating 
over $H_0$ (for the R98 data) or ${\cal M}_B$ (for the P99 data) and  
derive two-dimensional likelihood functions, $L(\Omega_0, \Omega_\Lambda)$
for the constant $\Lambda$ model and $L(\Omega_0, \alpha)$ for the 
spatially-flat time-variable $\Lambda$ case. These two-dimensional likelihood
functions are used to derive highest posterior density limits (see 
Ganga et al. 1997 and references therein) in the $(\Omega_0, \, 
\Omega_\Lambda)$ or $(\Omega_0, \, \alpha)$ planes.  In what follows we 
consider 1, 2, and 3 $\sigma$ confidence limits which include 68.27, 95.45, and
99.73\% of the area under the likelihood function.

When marginalizing over a parameter or deriving a limit from the likelihood
functions, we consider a number of different priors. We first consider a 
uniform prior in the parameter integrated over, set to zero outside the range 
considered for the parameter.

We also incorporate constraints from measurements 
which indicate $H_0 = 65 \pm 7\ {\rm km}\ {\rm s}^{-1}\ {\rm Mpc}^{-1}$ 
at 1 $\sigma$ (see, e.g., Biggs et al. 1999; Madore
et al. 1999), by using the prior
\begin{equation}
  p(H_0) = {1 \over \sqrt{2\pi}\{ 7\ {\rm km}\ {\rm s}^{-1}\ {\rm Mpc}^{-1}\}}
            {\rm exp}\bigg[ -{ \{ H_0 - 65\ {\rm km}\ {\rm s}^{-1}\ 
            {\rm Mpc}^{-1}\}^2 \over 2 \{ 7\ {\rm km}\ {\rm s}^{-1}\ 
            {\rm Mpc}^{-1}\}^2}\bigg] , 
\end{equation}
and from measurements which indicate $t_0 = 12 \pm 1.3$ Gyr at 1 $\sigma$ (i.e., 0.5 Gyr added to the globular cluster age estimate of Chaboyer 
et al. 1998)\footnote{
A more complete analysis would need to account for the uncertainty in this
(0.5 Gyr) numerical value. This would likely weaken the effect of this
prior.},
by using the prior
\begin{equation}
  p(t_0) = {1 \over \sqrt{2\pi}\{ 1.3\ {\rm Gyr} \}}
            {\rm exp}\bigg[ -{ \{ t_0(H_0, \Omega_0, \Omega_\Lambda) 
            - 12\ {\rm Gyr}
            \}^2 \over 2 \{ 1.3\ {\rm Gyr} \}^2}\bigg] , 
\end{equation}
with $\alpha$ replacing $\Omega_\Lambda$ in this expression in the 
spatially-flat time-variable $\Lambda$ case. Note that Figure 2 of Chaboyer et 
al. (1998) may be used to establish that a Gaussian prior of the form of 
eq. (4) is a good approximation to the shifted Monte Carlo globular cluster age 
distribution. See Ganga et al. (1997) for a discussion of Gaussian priors in a 
related context. This method of incorporating constraints from measurements
of $H_0$ and $t_0$ is not identical to the classical $H_0 t_0$ cosmological
test (see, e.g., $\S$13 of Peebles 1993). 

Finally, since $\Omega_0$ is a positive quantity, we also consider the 
non-informative prior (Berger 1985)\footnote{
We thank R. Gott for emphasizing this prior; a more complete discussion of
it may be found in Gott et al. (in preparation). I. Wasserman has noted
that such a prior is probably more appropriate for a parameter that sets
the scale for the problem, such as $H_0$ here (see $\S$B.2 of Drell et al.
1999; $\S$VII of Jaynes 1968 gives a more general discussion). It is however
still of interest to determine how the conclusions depend on the choice of 
prior.},
\begin{equation}
   p(\Omega_0) = {1 \over \Omega_0}. 
\end{equation}
Since $\Omega_0$ is bounded from below by the observed lower limit
on the baryon density parameter ($\sim 0.01 - 0.05$, depending on data used, 
see, e.g., Olive, Steigman, \& Walker 1999), this prior does not result in
an infinity.

\section{Results and Discussion}

Figure 1 shows the posterior probability density distribution function 
(PDF) confidence contours for the time-variable $\Lambda$ scalar field
model, derived using the three data sets discussed above. Figure 1$d$ 
shows that constraints from the three different data sets are quite 
consistent. At 2 $\sigma$ a large region of the parameter space
of these spatially-flat models (with a constant or time-variable $\Lambda$)
is consistent with the SNe Ia data. This data favors a smaller $\Omega_0$
as $\alpha$ is increased from zero.

A fluid with a time-independent equation of state $p = w \rho$, $w < 0$, has 
often been used to approximate the scalar field with 
potential $V(\phi) \propto \phi^{-\alpha}$
in the time-variable $\Lambda$ model. The solid lines in Figure 2$a$ show the
confidence contours for such a fluid model, derived using the P99 Fit C data.
These confidence contours are consistent with those shown in Figure 1
of Perlmutter et al. (1999b) (but see discussion below), Figure 10 of Wang 
et al. (1999), and Figure 4 of Efstathiou (1999). The dashed lines in Figure 
2$a$ show the exact scalar field model contours of Figure 1$c$, transformed 
using the relation between $w_\phi$ and $\alpha$ in the CDM and baryon 
dominated 
epoch (eq. [1]), which comes to an end just before the present. The two sets
of contours agree near $w \sim -1$, as they must since at $w = -1$ this is the 
flat-$\Lambda$ model and $\Lambda$ does behave exactly like a fluid with
a time-independent equation of state. However, the two sets of confidence 
contours differ significantly at larger $w$. Since the scalar field $w_\phi$
eventually switches over to $-1$ (in the scalar field dominated epoch, RP),
it is unclear what significance should be ascribed to this difference. We
stress, however, that since the time-independent equation of state fluid model 
is an approximation to the time-variable $\Lambda$ scalar field model, 
constraints on model-parameter values that are based on the fluid model 
approximation are only approximate (and possibly indicative).
In passing, we note that Perlmutter et al. (1999b)
use $w_{\rm eff}$ (eq. [2]) and not $w$ to parametrize the fluid model 
constraints. Figure 2$b$ shows contours of constant $w_{\rm eff}$ as a function
of $\alpha$ and $\Omega_0$ in the time-variable $\Lambda$ scalar field model.
In a large part of model parameter space $w_{\rm eff}$ is a sensitive function
of both $\alpha$ and $\Omega_0$ and hence is not the best parameter to use to
describe the time-variable $\Lambda$ scalar field model. 
 
Figure 3 shows the effects of incorporating constraints based on $H_0$
and $t_0$ measurements (eqs. [3, 4]). Panel $a)$ shows that adding the $H_0$
constraint does not significantly alter the contours derived from the
SNe Ia data alone. This is expected since the value of $H_0$ used here is
very close to the value that is indicated by the SNe Ia data (see R98).
However, incorporating the $t_0$ constraint, $t_0 = 12 \pm 1.3$ Gyr at 1 
$\sigma$, does significantly shift the contours, panel $b)$. This is 
because the SNe Ia data alone favor a higher $t_0$, $14.2 \pm 1.7$ Gyr
(R98), or $14.5 \pm 1.0\ (0.63/h)$ Gyr (P99).

Figure 4 shows constraints on the time-variable $\Lambda$ model, from the
three different SNe Ia data sets used in conjunction with the $H_0$ and
$t_0$ measurements (eqs. [3, 4]). The main effect of incorporating the 
$H_0$ and $t_0$ constraints is to increase the favored values of $\Omega_0$;
a weaker effect is the disfavoring of larger values of $\alpha$. Even this 
extended set of data does not tightly constrain model-parameter values.

Figure 5 shows the corresponding constraints on the constant $\Lambda$ 
model (from the SNe Ia, $H_0$, and $t_0$ measurements). Again, the major
effect of including the $H_0$ and $t_0$ data is to increase the favored 
values of $\Omega_0$. A weaker effect is that it reduces the odds against
reasonable open models (G\'orski et al. 1998), but not by a large factor.

Figures 6 and 7 show the effects of using the non-informative prior,
$1/\Omega_0$, of eq. (5). The major effect is a decrease in the favored 
values of $\Omega_0$. In the time-variable $\Lambda$ case there is also
a slight increase in the favored values of $\alpha$ (see Figure 6) while in
the constant $\Lambda$ case there is a mild reduction in the odds against
reasonable open models (see Figure 7). If the PDF was narrower (i.e., 
if the error bars on the data were smaller), changing from the flat to the 
non-informative prior would not result in as large a change in the confidence 
contours.

\section{Conclusion}

Recent SNe Ia data do favor models with a constant or time-variable $\Lambda$
over an open model without a $\Lambda$. However, this is not at a very high
level of statistical significance. Also, the incomplete understanding of a 
number of astrophysical effects and processes (evolution, intergalactic dust, 
etc.) means that these results are preliminary and not yet definitive. 

The constraints on the time-variable $\Lambda$ model derived here are based
on the exact scalar field model equations of motion, not on the widely used 
time-independent equation of state fluid approximation equations of motion.

\bigskip

We acknowledge the advice and assistance  of R. Gott, B. Kirshner, L. Krauss, 
V. Periwal,
S. Perlmutter, A. Riess, E. Sidky, M. Vogeley, and I. Wasserman, and are
specially indebted to J. Peebles and T. Souradeep. We acknowledge
helpful discussions with I. Waga who has also computed the PDF confidence
contours for the time-variable $\Lambda$ scalar field model, with results
consistent with those found in this paper. We thank the referee, I. Wasserman,
for a prompt and detailed report which helped us improve the manuscript.
We acknowledge support from NSF CAREER grant AST-9875031.

% Bibliography

\clearpage
\centerline{\bf Figure Captions}

\medskip
\noindent
Fig.~1.--
{\protect PDF confidence contours for the spatially-flat time-variable 
  $\Lambda$ scalar 
  field model, with potential $V(\phi) \propto \phi^{-\alpha}$, derived using
  the three SNe Ia data sets. The $\alpha = 0$ axis corresponds to the 
  spatially-flat time-independent $\Lambda$ case. Confidence contours in panels 
  $a)-c)$ run from $-2$ to $+3$ $\sigma$ starting from the lower left hand 
  corner of each panel. Panel $a)$ shows those derived from all the R98 SNe, 
  while $b)$ is for R98 SNe excluding the $z = 0.97$ one, and $c)$ is for the 
  P99 Fit C data set. Panel $d)$ compares the $\pm 2$ $\sigma$ limits from the 
  three data sets: all R98 SNe (long-dashed lines); R98 SNe excluding 
  the $z = 0.97$ one (short-dashed lines); and the P99 Fit C SNe (dotted 
  lines).} 

\medskip
\noindent
Fig.~2.--
{\protect $a)$ PDF confidence
  contours derived from the P99 Fit C SNe, for a spatially-flat fluid model 
  with equation of state 
  $p = w \rho$ (solid lines show contours from  $-3$ to $+3$ $\sigma$ starting 
  from the lower left hand corner) and for the spatially-flat
  time-variable $\Lambda$ scalar 
  field model with potential $V(\phi) \propto \phi^{-\alpha}$ which behaves
  like a fluid model with equation of state $p_\phi = w_\phi \rho_\phi$ in
  the CDM and baryon dominated epoch (dotted lines 
  show contours from  $-2$ to $+3$ $\sigma$ starting from the lower left hand   
  corner). The scalar field model contours were derived by using eq. (1)
  to transform the vertical axis of Figure 1$c$. The fluid model likelihood
  function was computed for the full range of $w$ and $\Omega_0$ shown.
  The dot-dashed lines bound the $w_\phi-\Omega_0$ region that corresponds
  to the $\alpha-\Omega_0$ region of Figure 1$c$ over which the scalar field
  model likelihood was computed. 
  $b)$ Contours of constant $w_{\rm eff}$ (eq. [2]) in the $\alpha-\Omega_0$
  plane for the time-variable $\Lambda$ scalar field model with potential 
  $V(\phi) \propto \phi^{-\alpha}$. Starting at the lower left hand corner, the 
  contours correspond to $w_{\rm eff} = -0.9, -0.8, -0.7, -0.6, -0.5, -0.4,
  -0.35, -0.3,$ and $-0.25$.}

\medskip
\noindent
Fig.~3.--
{\protect PDF confidence
  contours derived from the P99 Fit C SNe, for the spatially-flat 
  time-variable $\Lambda$ 
  scalar field model with potential $V(\phi) \propto \phi^{-\alpha}$.
  The dotted lines in panels $a)-c)$ show the $-2$ to $+3$ $\sigma$ contours 
  of Figure 1$c$. The solid lines in these panels show the corresponding 
  contours derived from the SNe data in conjunction with: $a)$ $H_0$ 
  measurements
  (using eq. [3]); $b)$ $t_0$ measurements (using eq. [4]); and $c)$
  $H_0$ and $t_0$ measurements (using eqs. [3, 4]). Panel $d)$ compares the
  $\pm 2$ $\sigma$ confidence contours from the SNe data in conjunction with:
  the $H_0$ constraint (long-dashed lines); the $t_0$ constraint (short-dashed
  lines); and both constraints (dotted lines).} 

\medskip
\noindent
Fig.~4.-- 
{\protect PDF confidence contours for the spatially-flat time-variable 
  $\Lambda$ scalar field model, with 
  potential $V(\phi) \propto \phi^{-\alpha}$, derived using
  the three SNe Ia data sets. Panel $a)$ shows those derived from all
  the R98 SNe, while $b)$ is from the R98 SNe excluding the $z = 0.97$ one, 
  and $c)$ is for the P99 Fit C data. Solid (dotted) lines in $a)-c)$ are 
  the 1, 2, and 3 $\sigma$ contours 
  from the SNe data with (without) the $H_0$ and $t_0$ constraints (eqs. [3,
  4]); the dotted lines here are the solid lines in Figure 1$a-c$.
  Panel $d)$ compares the $\pm 2$ $\sigma$ limits from the $H_0$ and 
  $t_0$ constraints used in conjunction with: all R98 SNe (long-dashed lines); 
  R98 SNe excluding the $z = 0.97$ one (short-dashed lines); and the P99 Fit C 
  SNe (dotted lines).}

\medskip
\noindent
Fig.~5.--
{\protect PDF confidence contours for the time-independent $\Lambda$ model, 
  derived using the three SNe Ia data sets. Panel $a)$ shows those derived from 
  all the R98 SNe, while $b)$ is from the R98 SNe excluding the $z = 0.97$ one, 
  and $c)$ is for the P99 Fit C SNe. Solid (dotted) lines in $a)-c)$ are the 
  1, 2, and 3 $\sigma$ contours from the SNe data with (without) the $H_0$ and 
  $t_0$ constraints (eqs. [3, 4]). Panel $d)$ compares the $\pm 2$ $\sigma$ 
  limits from the $H_0$ and $t_0$ constraints used in conjunction with: all R98 
  SNe (long-dashed lines); R98 SNe excluding the $z = 0.97$ one (short-dashed 
  lines); and the P99 Fit C SNe (dotted lines). In all panels, models with 
  parameter values in the upper left hand corner region bounded by the diagonal 
  dot-dashed curve do not have 
  a big bang. The horizontal dot-dashed line demarcates models with a zero
  $\Lambda$ and the diagonal dot-dashed line running from $\Omega_\Lambda = 1$
  to $\Omega_0 = 2$ corresponds to spatially-flat models.}

\medskip
\noindent
Fig.~6.--
{\protect PDF confidence
  contours (1, 2, and 3 $\sigma$) derived from the P99 Fit C SNe, for the 
  spatially-flat time-variable $\Lambda$ scalar field model. Panel 
  $a)$($b)$) ignores
  (accounts for) the $H_0$ and $t_0$ constraints (eqs. [3, 4]). The solid
  (dotted) lines use the non-informative $1/\Omega_0$ (flat) prior.
  The dotted lines in $a)$($b)$) are the same as the solid lines in Figure
  1$c$ (4$c$).}

\medskip
\noindent
Fig.~7.--
{\protect PDF confidence
  contours (1, 2, and 3 $\sigma$) derived from the R98 SNe excluding the
  $z = 0.97$ one, for the time-independent $\Lambda$ model. Panel $a)$($b)$) 
  ignores (accounts for) the $H_0$ and $t_0$ constraints (eqs. [3, 4]). The 
  solid (dotted) lines use the non-informative $1/\Omega_0$ (flat) 
  prior. The dotted lines in $a)$($b)$) are the same as the dotted (solid) 
  lines in Figure 5$b$. The dot-dashed lines are described in the caption of 
  Figure 5. In the non-informative prior cases the likelihood function
  is computed down to $\Omega_0 = 0.01$.}

\clearpage

\begin{figure}
\resizebox{\textwidth}{!}{\includegraphics{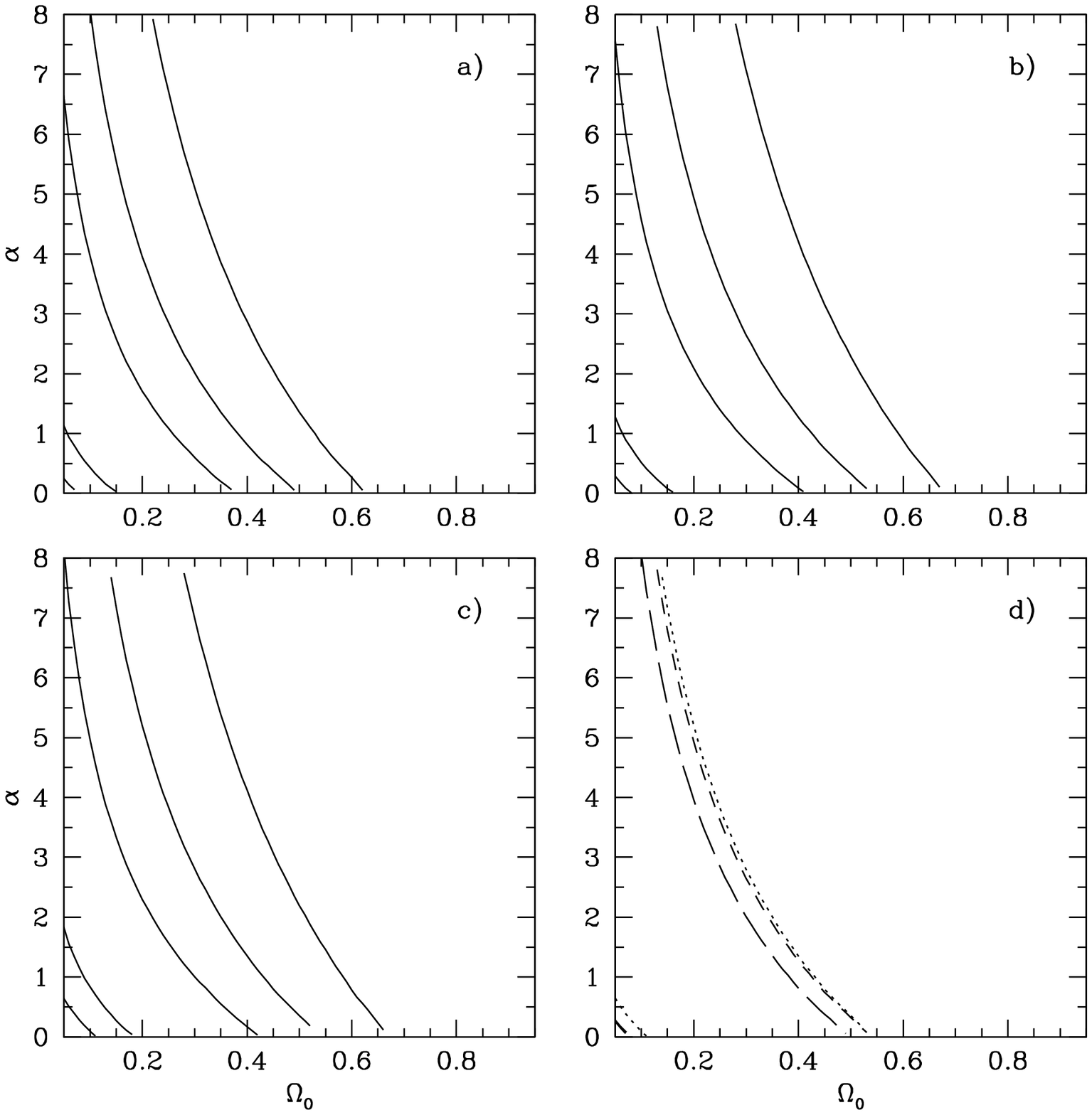}}
Figure 1
\end{figure}
\clearpage

\begin{figure}
\resizebox{\textwidth}{!}{\includegraphics{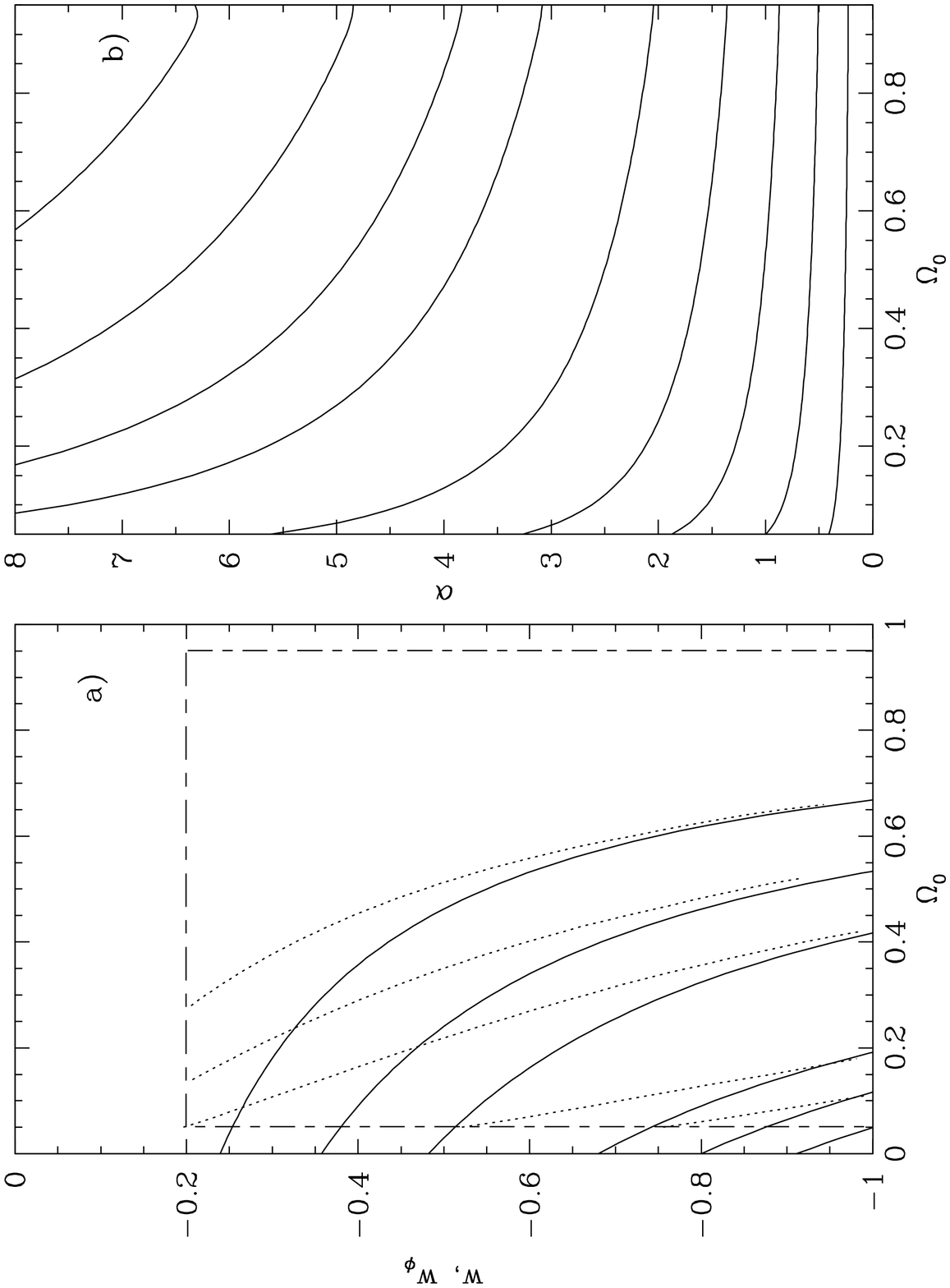}}
Figure 2
\end{figure}
\clearpage

\begin{figure}
\resizebox{\textwidth}{!}{\includegraphics{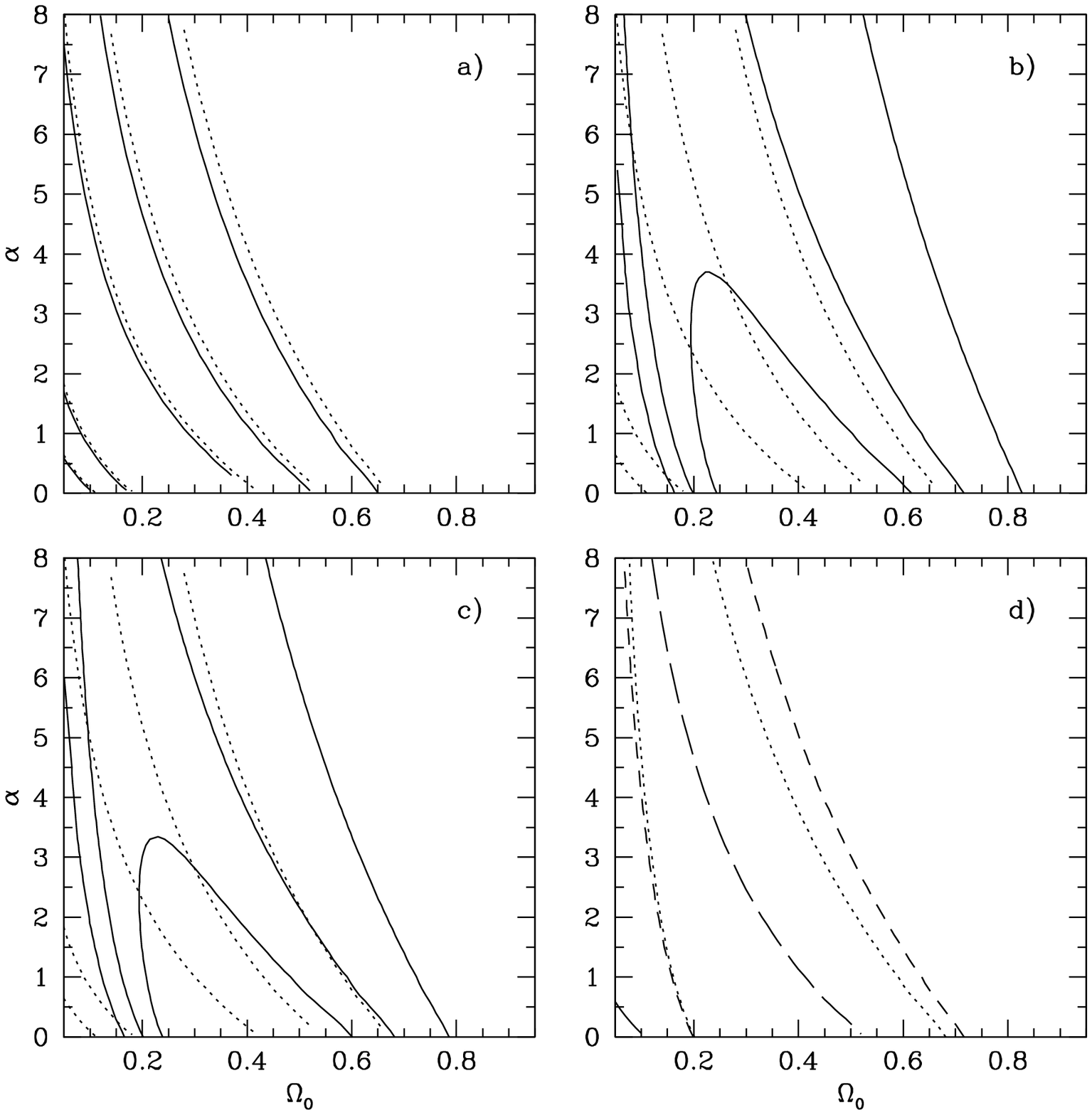}}
Figure 3
\end{figure}
\clearpage

\begin{figure}
\resizebox{\textwidth}{!}{\includegraphics{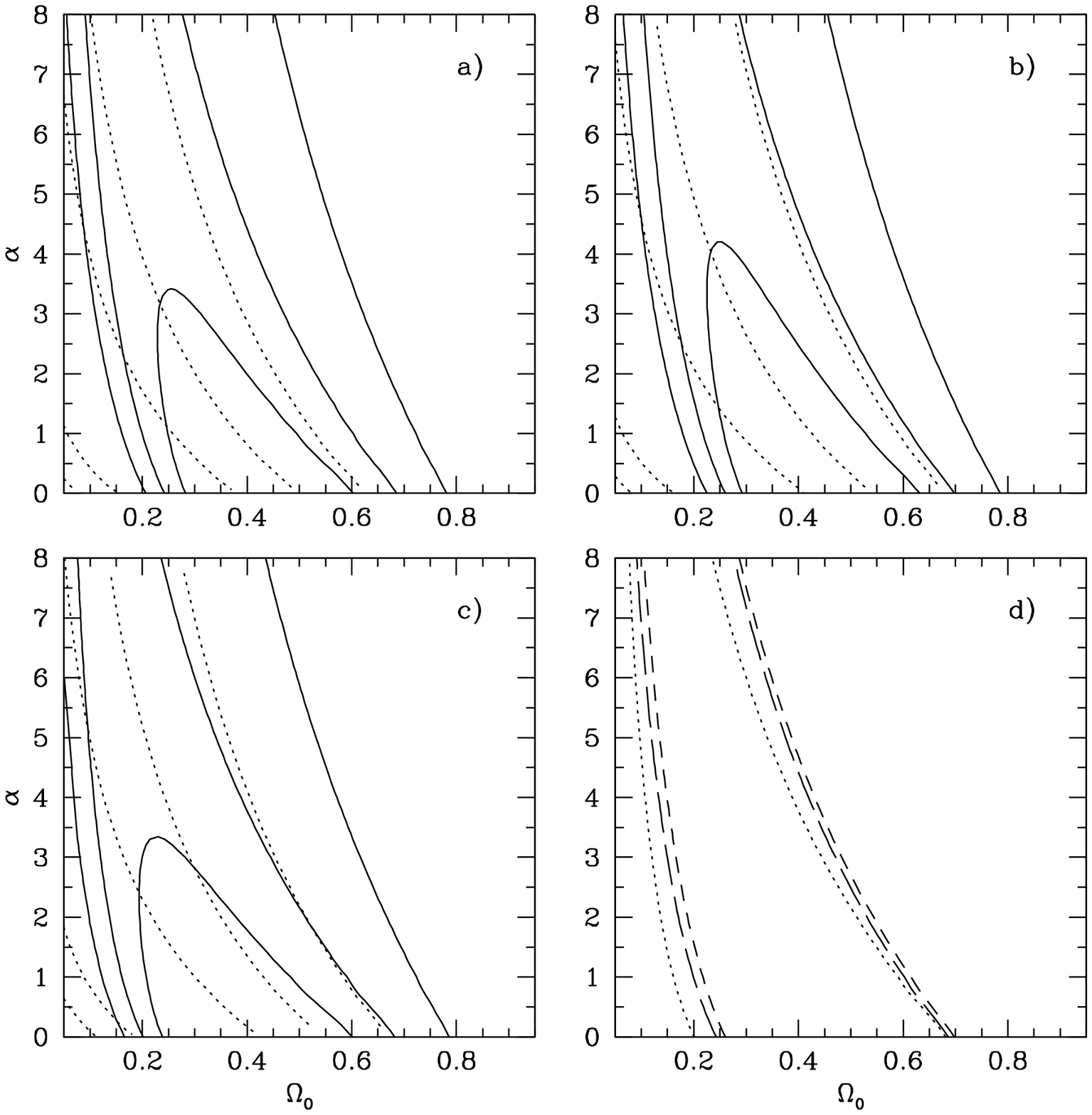}}
Figure 4
\end{figure}
\clearpage

\begin{figure}
\resizebox{\textwidth}{!}{\includegraphics{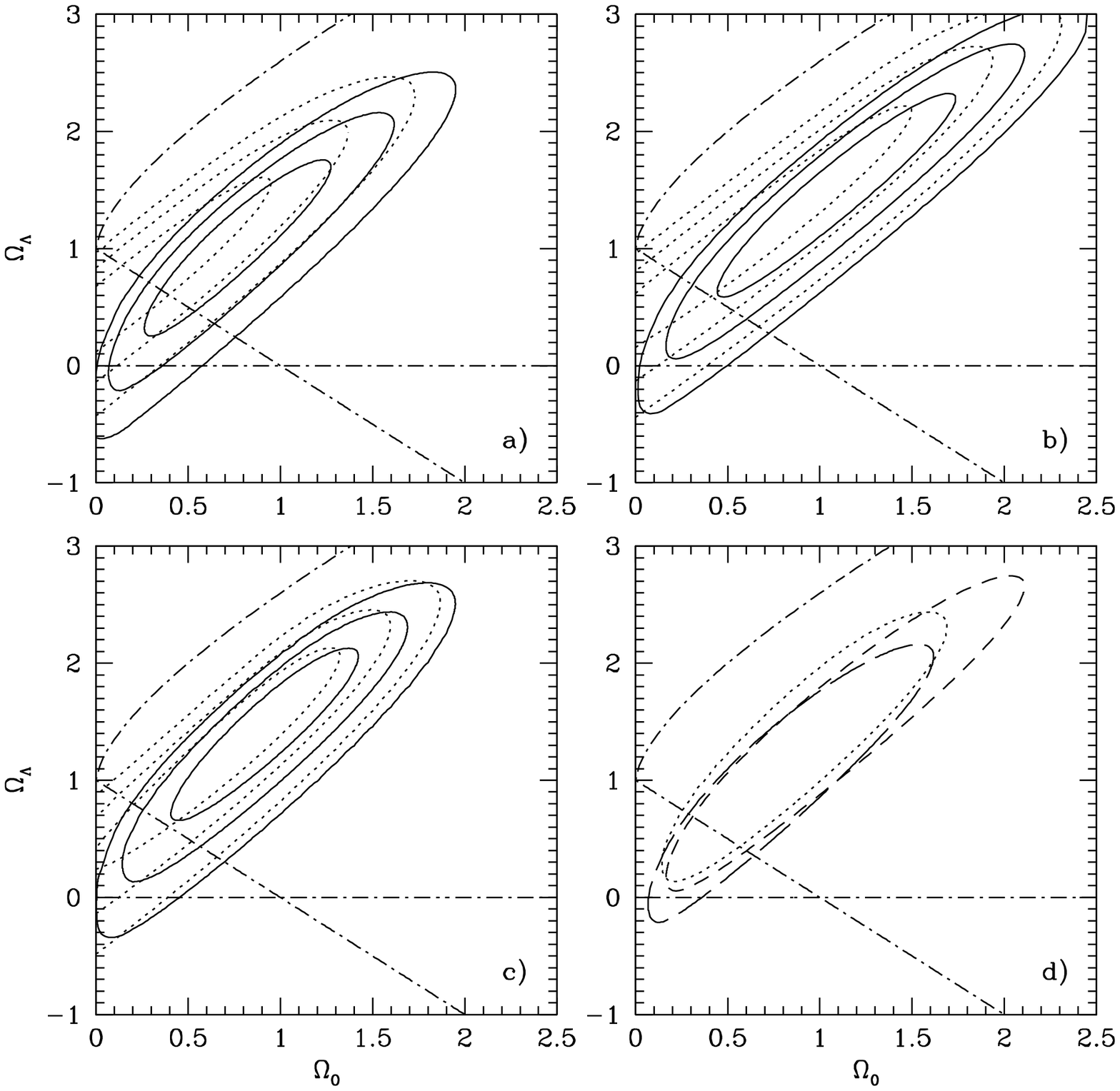}}
Figure 5
\end{figure}
\clearpage

\begin{figure}
\resizebox{\textwidth}{!}{\includegraphics{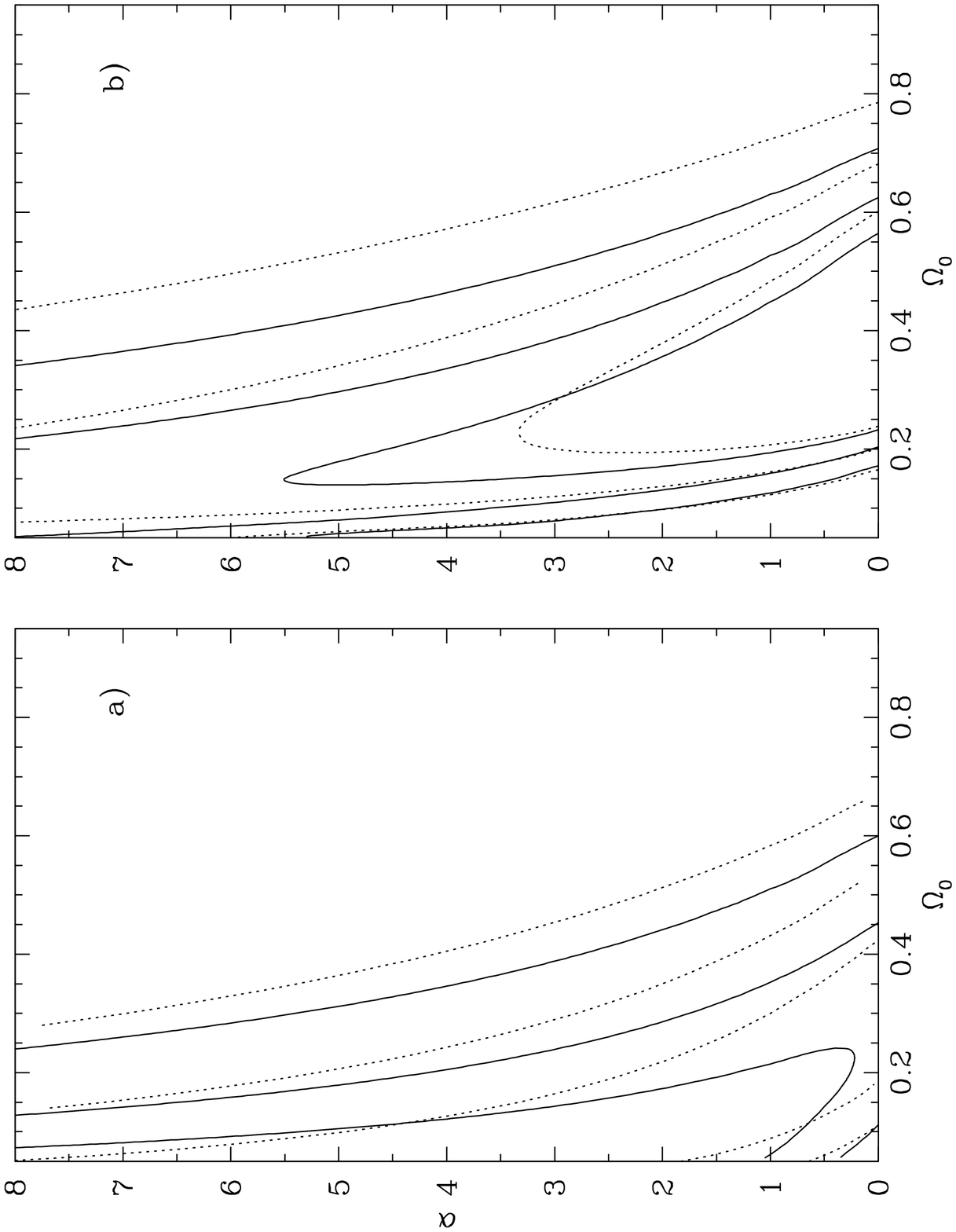}}
Figure 6
\end{figure}
\clearpage

\begin{figure}
\resizebox{\textwidth}{!}{\includegraphics{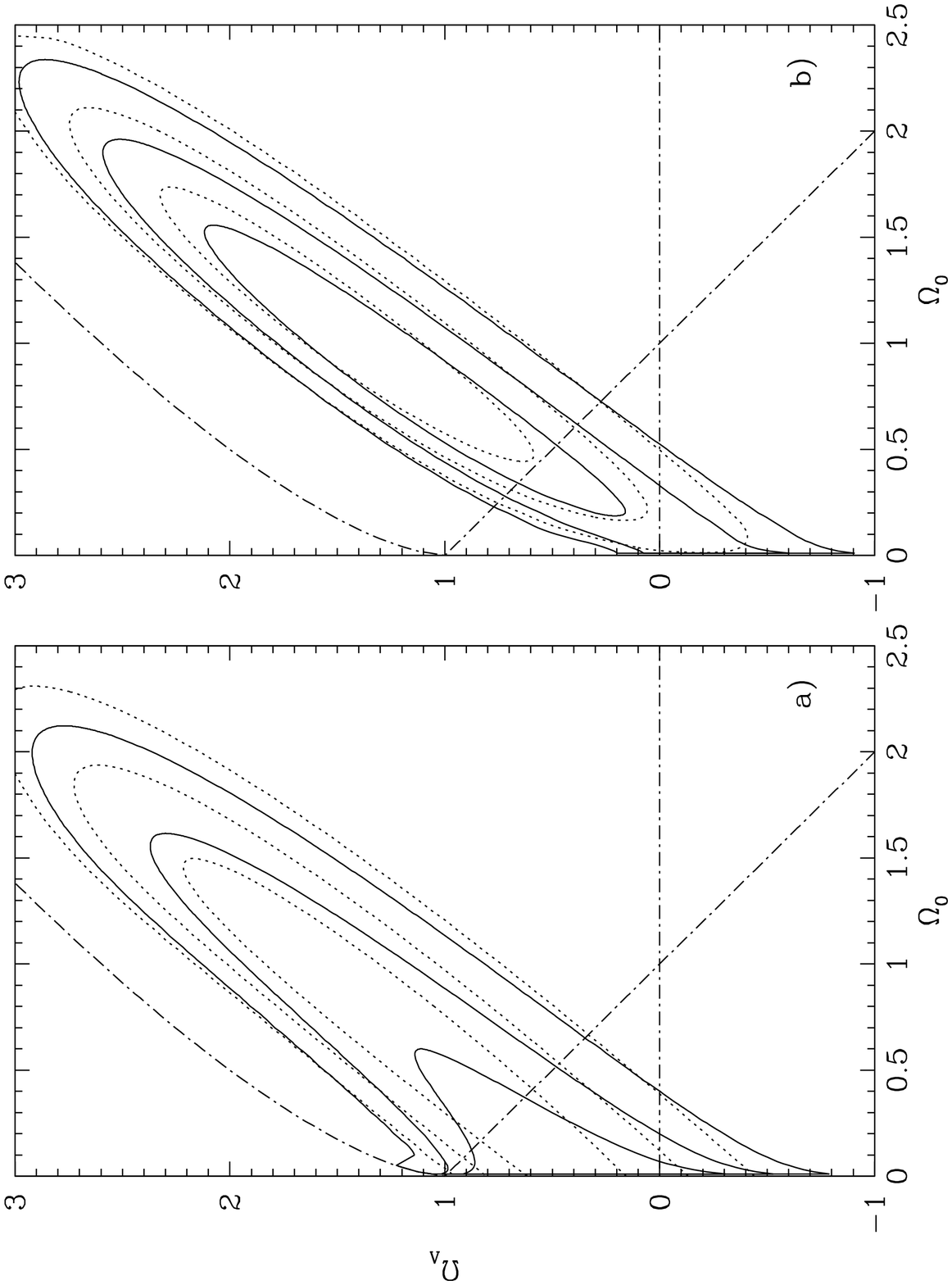}}
Figure 7
\end{figure}
\clearpage

\end{document}